\documentclass[]{aa}
\usepackage{graphics}
\begin{document}
\thesaurus{06(08.16.5; 08.02.3; 08.03.4; 09.10.1; 02.16.2; 02.19.2)}

\title{ Circumstellar disks around Herbig Ae/Be stars: polarization, outflows
and binary orbits}
\author {G. Maheswar, P. Manoj  and H. C. Bhatt}
\institute {Indian Institute of Astrophysics, Bangalore 560 034,
India}
\offprints {G. Maheswar}
\mail {maheswar@iiap.ernet.in,\\
manoj@iiap.ernet.in, hcbhatt@iiap.ernet.in}

\date{}
\titlerunning{Disk orientation in Herbig Ae/Be stars}

\maketitle
\begin{abstract}
The geometrical relationship between the distribution of circumstellar matter, 
observed optical linear polarization, outflows and binary orbital plane
in Herbig Ae/Be stars is investigated.
Optical linear polarization measurements carried out for a number of 
Herbig Ae/Be stars that are either
known to be in binary systems and/or have bipolar jets are presented
in this paper. Available information on the position angles of polarization,
outflows and binary companions for Herbig Ae/Be stars is compiled and
analysed for any possible correlations. In $\approx 85\%$ of the sources 
the outflow position angle is within $30^{\circ}$ of being parallel or 
perpendicular to the polarization
position angle. In $\approx 81\%$ of the sources
the binary position angle is within $30^{\circ}$ of being parallel
or perpendicular to the polarization position
angle. Out of 15 sources with bipolar outflows, 10 sources have the binary 
position 
angle  within $30^{\circ}$ of being perpendicular to the outflow 
position angle. These results favour those binary formation mechanisms 
in which the binary components and the disks around individual stars or 
circumbinary disks are coplanar.\\


\end{abstract}

\section{Introduction}

Herbig Ae/Be stars are pre-main sequence (PMS) objects of intermediate
mass ($2 \le M/M_{\odot} \le 8$). In the original survey (Herbig 1960)
these objects were defined as A and B stars, located in regions of known
star formation, with emission in the Balmer lines of hydrogen and
associated with optical reflection or emission nebulosity. Herbig (1960)
listed 26 objects belonging to this class. The catalogue was later expanded
by Finkenzeller \& Mundt (1984) to include 57 stars. The most recent
catalogue by Th\'{e} et~al. (1994), adopting a more extended
definition, lists
109 Herbig Ae/Be stars and a number of candidates that include stars with
later spectral types (G0 or earlier) and those found relatively isolated
from star forming clouds. The PMS nature of Herbig Ae/Be stars is now
well established, based on their position in the HR diagram and comparison
with theoretical evolutionary tracks (Strom et~al. 1972; Cohen \& Kuhi
1979; van den Ancker et~al. 1998; Palla \& Stahler 1991).
Infrared, submillimetre and millimetre measurements have shown that Herbig
Ae/Be stars are associated with significant amounts of circumstellar
dust emitting excess radiation, over that produced by stellar photosphere,
at these wavelengths (eg., Rydgren et~al. 1976;  Cohen \& Kuhi
1979; Bertout et~al. 1988; Strom et~al. 1988; Beckwith et~al.
1990; Weintraub Sadnell \& Duncan 1989; Adams et~al. 1990;
Hillenbrand et~al. 1992). The existence of circumstellar dust is also
supported by the relatively large values of intrinsic polarization
observed for these objects (eg., Breger 1974; Garrison \& Anderson
1978; Vrba et~al. 1979; Jain et~al. 1990; Jain \& Bhatt 1995; Yudin \&
Evans 1998) which is generally ascribed to the presence of circumstellar dust
grains (eg., Bastien 1987). \\

While the existence of circumstellar dust around Herbig Ae/Be stars is
well established, the geometrical distribution of the dust is not yet
fully clear. Hillenbrand et~al. (1992), from an analysis of the spectral
energy distributions (SED) of 47 Herbig Ae/Be stars, classified these
objects into three groups. The infrared  SED of the Group I objects
($\lambda F_{\lambda} \sim \lambda^{-4/3}$) could be
explained by invoking a geometrically thin, optically thick circumstellar
accretion disk with an optically thin inner region to account for the
observed inflections in their near-infrared spectra. Group II objects,
with flat or rising infrared spectra, consist of a star or star/disk
system surrounded by gas and dust that is not confined to a disk. Group
III objects have small or no infrared excess, similar to classical Be
stars, and the small excesses can be ascribed to free-free emission in a
gaseous envelope. Berrilli et~al. (1992), on the other hand, produced
models with spherically symmetric dust envelopes. Natta et~al. (1993)
proposed models with at least three components that contribute to the
observed infrared emission: the central star, a circumstellar disk, and an
extended, almost spherically symmetric, envelope. Models with symmetrical
envelopes generally lead to much higher visual extinctions toward the central
star than are observed. \\

Polarimetric measurements provide an important tool to study the nature
and geometry of the circumstellar material. The integrated light from a
Herbig Ae/Be object can have intrinsic polarization (in addition to the
interstellar component) only if the distribution of the scattering material
in its circumstellar regions is not spherically symmetric. Circumstellar
dust distributed in a disk can cause relatively large polarization, the
degree of polarization depending on the amount of scattering dust, degree
of flattening of the disk and its orientation with respect to the
observer's line of sight to the star. If the disk optical depth is small
($\tau \le 0.3$) and polarization is produced by single scattering, then
the position angle of the $E$ vector of the linearly polarized light is
perpendicular to the disk, while for optically thick disks the observed
polarization may be dominated by optically thin scattering dust
distributed perpendicularly to the disk (for example in bipolar jets and
outflows) resulting in a polarization position angle that is parallel to
the disk plane (eg., Brown \& McLean 1977; Elsasser \& Staude 1978).
Polarization vectors perpendicular to the disk plane can also be obtained
for scattering off the surface of optically thick disks, while polarization
vectors parallel to the disk plane require the addition of an extensive
circumstellar envelope (see eg., Whitney \& Hartmann 1992, 1993).
Some Herbig Ae/Be stars, similarly to their lower mass counterparts, the
T Tauri stars, have also been found to exhibit bipolar optical jets and
molecular outflows (eg., Canto et~al. 1984; Strom et~al. 1986; Corcoran \&
Ray 1998) which are believed to be directed perpendicular to the
circumstellar disks (eg. Konigl, 1982). One may therefore expect a
correlation between the position angle of polarization and the direction
of the jets and outflows from the Herbig Ae/Be stars. In a sample of 23 T
Tauri stars and other young stellar objects, including 8 Herbig Ae/Be
stars, Bastien (1987) found that for about 50 to 60 \% of the sources the
directions of the outflow and of the polarization are within $30^{\circ}$ of
being perpendicular to each other. If only the 8 Herbig Ae/Be stars in
their sample are considered, then the correlation is not clear. In recent
years more extensive polarization measurements and observations of jets
and outflows from  more Herbig Ae/Be stars have become available. A study
of the correlation between the polarization and outflow directions can now
be performed on a larger sample of Herbig Ae/Be stars.\\

A relatively large fraction (as compared with that for the main-sequence
field stars) of Herbig Ae/Be stars have been found to have binary companions
(eg., Leinert et~al. 1997). Are the circumstellar (or circumbinary) disks
in the binary Herbig Ae/Be systems coplanar with the binary orbital plane?
Their relative orientations may depend on the binary formation process.
For example, fragmentation of  rotating protostellar clouds during
collapse (eg. Boss 1988) to form binary systems will favour the formation
of disks that are coplanar with the orbital plane. On the other
hand, formation of the binary system by capture of independent young stellar
or protostellar objects (eg., Larson 1990) will not favour any
such alignment. An observational study of the relative orientations of
the polarization position angles, bipolar outflows and binary orbital
planes may therefore be useful in discriminating between the different
models for the origin of binary systems. \\

In this paper we present the results of our measurements of optical linear
polarization for a number of Herbig Ae/Be stars that are either known to
be in binary systems and/or have bipolar jets. Available information on
the position angles of polarization, outflows and binary companions for
Herbig Ae/Be stars is compiled and analysed for any possible correlations.
Our polarimetric observations are presented in Section~\ref{obs}. Data from the
literature is compiled in Section~\ref{dat}. Various correlations are presented in section~\ref{rel} 
and discussed in Section~\ref{dis}. Conclusions are summarized in Section~\ref{con}.\\

\section{Observations}
\label{obs}

\begin{table}
\caption[]{Polarization of the observed Herbig Ae/Be stars.}
\begin{tabular}{l c c c c c c}
\hline\hline
Object  & Date of &  P &$\theta_{p}$ &$\epsilon_{p}$&$\epsilon_{\theta}$ \\
        & observation&($\%$) &($^{\circ}$)&($\%$)&($^{\circ}$)   \\
\hline
HD 35187 & 03 Mar 00 &  0.18  &128&0.07 &8   \\
GU CMa   & 10 Mar 99 &  1.19  &28 &0.07 &2   \\
NX Pup   & 02 Mar 00 &  1.23  &38 &0.23 &6   \\
Her 4636 & 01 Mar 00 &  0.77  &157&0.12 &6   \\
HD 141569& 29 Apr 00 &  0.79  &82 &0.05 &1   \\
HD 144432& 30 Apr 00 &  0.47  &20 &0.06 &4   \\

\hline
\end{tabular}
\label{tab1}
\end{table}

Optical linear polarization measurements of 6 Herbig Ae/Be stars were
made with a fast star and sky chopping polarimeter (Jain \& Srinivasulu
1991) coupled at the $f/13$ Cassegrain focus of the $1m$ telescope at
Vainu Bappu Observatory, Kavalur of the Indian Institute of Astrophysics.
A dry-ice cooled R943-02 Hamamatsu photomultiplier tube was used as the
detector. All measurements were made in the $V$ band with an aperture of
$15 arcsec$. Observations were made during the period of 1999-2000. The
instrumental polarization was determined by observing unpolarized standard
stars from Serkowski (1974). It was found to be $\sim 0.1 \%$, and has
been subtracted vectorially from the observed polarization of the
programme stars. The zero of the polarization position angle was
determined by observing polarized standard stars from Hsu \& Breger
(1982). The position angle is measured from the celestial north,
increasing eastward. The Herbig Ae/Be stars selected for observations were
taken from the Th\'{e} et~al. (1994) catalogue. HD 35187 (Dunkin et al. 1998) and
Her 4636 ( Williams et~al. 1977) satisfy all the criteria of being Herbig Ae/Be stars 
and hence we have made polarimetric observations
of these objects and included them in our sample. The selected objects are
either known to have outflows and/or are in binary systems. Results
of our polarimetric measurements are given in Table 1.
Columns in Table ~\ref{tab1} give respectively, (1) identification of the star,
(2) date of observation, (3) degree of polarization, (4) polarization
position angle, (5-6) $1\sigma$ probable errors in measurements of polarization
and the position angle.\\

\begin{table*}[!]
\caption[]{Data on binarity and outflows in Herbig Ae/Be stars}
\begin{tabular}{l c c c c c c c l}
\hline\hline
Object  &Dis. &Sp.typ &$\theta_{b}$&Angular&Linear&Mode of&$\theta_{o}$&Ref\\
        &     &       &            &Separation&Separation&detection      &            &   \\
        &(pc) &      &($^{\circ})$&($arcsec$)&(AU)&&($^{\circ}$)&\\
\hline
V633 Cas  &600  &A5 &3   &5.5&3300 &Near Infrared (NIR)  &160  &1,11 \\
V376 Cas  &600  &F0 &-   &-  &-    &-                    &120  &3\\
XY Per    &160  &B6 &255 &1.2&192  &NIR                  &-    &2\\ 
V892 Tau  &140  &A6 &23  &4.1&570  &NIR                  &-    &1\\
\vspace{0.2cm}
UX Ori    &460  &A2 &257 &0.02&10   &Optical              &-    &3\\
HD 35187  &150  &A2 &185 &1.3&195  &Optical              &-    &10\\
CO Ori    &460  &F8 &280 &2.0&920  &Optical              &-    &3\\
HK Ori    &460  &A5 &42  &0.3&138  &NIR                  &160  &1,12\\
T Ori     &460  &B9 &73  &7.7&3540 &NIR                  &-    &1\\
\vspace{0.2cm}
V380 Ori  &460  &B9 &204 &0.15&71   &NIR                  &56,149&1,11\\
LkH$\alpha$ 208
          &1000 &B7 &114 &0.12&120 &NIR                  &0    &1,13\\
GGD 18    &1600 &B2 &254 &5.8&9280 &NIR                  &150  &22,23\\
R Mon     &800  &B0 &331 &0.7&560  &NIR                  &0     &4,11\\
Gu CMa    &1150 &B1 &189 &0.7&800  &Optical              &-    &5\\
\vspace{0.2cm}
Z CMa     &1150 &B5 &123 &0.1&115  &NIR                  &60   &1,14\\
HD 53367  &240  &B0 &298 &0.7&167  &Optical              &-    &5\\
NX Pup    &500  &A0 &62  &0.13&65   &NIR &-    &6\\
HD 76534  &400  &B3 &304 &2.0&800  &Optical &-    &5\\
Her 4636  &600  &B2 &34  &3.3&1980 &Optical&148  &9,20\\
\vspace{0.2cm}
HD 141569 &100  &A0 &312 &6.8&680  &NIR    &-    &2 \\
HD 144432 &250  &A7 &354 &1.2&300  &Optical &-    &5\\
HR 5999   &200  &A0 &115 &1.4&280  &NIR  &-    &1\\
HD 150193 &150  &A2 &236 &1.1&165  &NIR  &-    &2\\
KK Oph    &160  &A6 &247 &1.5&240  &NIR  &-    &1\\
\vspace{0.2cm}
HD 163296 &120  &A0 &-   &-  &-    &-       &8    &15\\
AS 310    &2500 &B0 &153 &4.2&11000&NIR  &55   &7,12\\
R CrA     &130  &F0 &-   &-  &-    &-       &130  &13 \\
T CrA     &130  &F5 &275 &0.14&18   &Optical&133  &8,13\\
Par 21    &400  &A5 &-   &-   &-    &-       &155  &21\\
\vspace{0.2cm}
V1685 Cyg &1000 &B2 &23  &0.14&141  &Optical &-    &5\\
MWC 349   &1200 &B[e]&280&2.4 &2880 &Optical &10   &25\\
LkH$\alpha$ 234
          &1000 &B3 &315 &2.7&2700 &NIR  &226,252  &1,18,21\\
PV Cep    &600  &A5 &-   &-  &-    &-       &348  &16\\
HD 200775 &400  &B2 &164 &2.3&900  &NIR  &70   &2,17\\
\vspace{0.2cm}
AS 477    &900  &A0 &43  &4.2&3762 &NIR  &112  &2,12\\
LkH$\alpha$ 233
          &800  &A5 &-   &-  &-    &-       &50,90&13\\
HD 216629 &725  &B2 &147 &7.0&5046 &NIR  &-    &1\\
MWC 1080  &2500 &B0 &86  &4.7&11725&NIR  &60   &1,19\\
\hline \\
\end{tabular}
\label{tab2}

\underline{\textit{Notes:}}\\

\textit{V380 Ori and LkH$\alpha$233 have two outflow components.}\\
\textit{LkH$\alpha$234 shows an inner infrared jet at position angle $226^{\circ}$ and an outer
optical jet at a position angle $252^{\circ}$ (Cabrit et al.~1997).}\\
\textit{MWC 1080 is a triple systems consisting of a close unresolved spectroscopic binary and a tertiary star. Position angle of the tertiary is given here.}\\

\underline{References}\\

1) Leinert et~al. 1997; 2)Pirzkal et~al. 1997; 3) Bertout et~al. 1999; 4) Close 1997; 
5) Dommanget 1994; 6) Brandner et~al. 1995; 7) Ageorges et~al. 1997; 8) Bailey J. 1998; 
9) Chelli et~al. 1995; 10) Dunkin et~al. 1998; 11) Strom et~al. 1986; 12) Goodrich 1993; 
13) Bastien 1987; 14) Poetzel et~al. 1989; 15) Devine et~al. 2000; 16) Reipurth et~al. 1997; 
17) Watt 1986; 18) Ray et~al. 1990; 19) Poetzel et~al. 1992; 20) White 1993; 21) Cabrit et al. 1997; 22) 
Lenzen et al. 1984; 23) Lada and Gautier 1982; 24) Meyer et al. 2001; 25) White and Becker 1985.
\end{table*}

\begin{table*}[!]
\caption[]{Data on polarization in Herbig Ae/Be stars}
\begin{tabular}{l c c c c c c c c c c c l r}
\hline\hline
Object  &$\langle{P_{\star}}\rangle$&$\langle\theta_{\star}\rangle$&P$_{ism}$&
$\epsilon_{p_{ism}}$&$\theta_{ism}$&$\epsilon_{\theta_{ism}}$
&$\langle{P_{i}}\rangle$&$\langle\theta_{i}\rangle$&n&$\sigma_{p}$&$\sigma_{\theta}$&Aperture&Ref\\ 
   & (\%)      &($^{\circ}$)    &(\%)&(\%)&($^{\circ}$)&($^{\circ}$) &(\%)      &($^{\circ}$)&  &(\%)
&($^{\circ}$)  &($arcsec$) &  \\
(1)&(2)&(3)&(4)&(5)&(6)&(7)&(8)&(9)&(10)&(11)&(12)&(13)&(14)\\
\hline
V633 Cas   &1.67 &25  &1.24 &0.26&72  &6 &2.30  &9   &3 &0.26&10 &10,15,8.3           &1,2       \\
V376 Cas   &22.0 &26  &1.22 &0.26&71  &6 &23.5  &24  &7 &1.65&3  &10,15,8.3           &1,2       \\
XY Per     &1.58 &126 &0.15 &0.39&140 &52&1.45  &124 &7 &0.07&3  &10,15               &3,4       \\
V892 Tau   &4.72 &3   &-    &-   &-   &- &4.72  &3   &1 &0.29&2  &12                  &5         \\
\vspace{0.2cm}
UX Ori     &1.22 &99  &0.26 &0.12&123 &13&1.07  &94  &10&0.21&7  &20,10               &6,4       \\
HD 35187   &0.18 &128 &0.51 &0.29&61  &17&0.65  &145 &1 &0.07&8  &15                  &15        \\
CO Ori     &2.28 &164 &0.17 &0.39&97  &52&2.39  &165 &9 &0.28&9  &4.3,10              &10,4      \\
HK Ori     &1.10 &159 &0.08 &0.35&89  &52&1.17  &161 &5 &0.18&6  &10,15,-,13          &1,3,7,8,9 \\
T Ori      &0.32 &94  &0.15 &0.16&148 &32&0.28  &100 &11&0.10&40 &10,-,13             &3,7,8,9,4 \\
\vspace{0.2cm}
V380 Ori   &0.92 &88  &0.16 &0.19&72  &35&0.80  &91  &6 &0.29&7  &15,-,14.3,20        &3,7,8,11,6\\
LkH$\alpha$ 208
           &2.28 &6   &1.46 &0.38&172 &8 &1.24  &25  &4 &0.49&11 &10,15,13,-          &1,9,7,8   \\
GGD 18     &3.60 &81  &-    &-   &-   &- &3.60  &81  &1 &0.50&2  &                    &24        \\
R Mon      &13.0 &92  &0.70 &0.34&167 &14&13.5  &91  &10&1.96&8  &26,8,13,15,10       &12,1,8\\
Gu CMa     &1.25 &21  &0.47 &0.36&146 &22&1.46  &29  &4 &0.07&4  &10,15               &13,14,3,15\\
\vspace{0.2cm}
Z CMa      &0.80 &125 &0.36 &0.32&135 &26&0.67  &173 &5 &0.26&23 &10,-,20,15          &3,7,8,16  \\
HD 53367   &0.53 &36  &0.14 &0.28&152 &52&0.62  &41  &2 &0.04&1  &10,15               &3,8       \\
NX Pup     &1.10 &47  &0.28 &0.13&38  &13&0.82  &50  &2 &0.21&10 &20,10               &15,16     \\
HD 76534   &0.61 &126 &0.15 &0.38&88  &52&0.59  &133 &1 &0.29&11 &15                  &3         \\
Her 4636   &0.80 &159 &0.12 &0.06&86  &14&0.90  &162 &2 &0.05&1  &-,15                &22,15     \\
\vspace{0.2cm}
HD 141569  &0.57 &88  &1.00 &0.71&86  &20&0.43  &175 &10&0.71&20 &15                  &4,15      \\
HD 144432  &0.30 &30  &0.24 &2.38&36  &52&0.19  &172 &5 &2.39&52 &15                  &4,15      \\
HR 5999    &0.52 &177 &0.96 &0.62&17  &19&0.68  &121 &2 &0.62&19 &-                   &29        \\
HD 150193  &4.91 &59  &-    &-   &-   &- &4.91  &59  &6 &0.51&9  &15,12,20            &3,4,5     \\
KK Oph     &5.10 &180 &0.13 &0.26&179 &52&4.92  &180 &10&1.33&8  &20,10               &16,4      \\
\vspace{0.2cm}
HD 163296  &0.21 &59  &0.70 &0.45&162 &19&0.76  &70  &13&0.45&19 &15,13,27            &3,17,18   \\
R CrA      &7.6  &189 &0.02 &0.09&31  &52&7.6   &9   &1 &0.50&5  &5                   &21        \\
T CrA      &4.59 &181 &0.02 &0.09&31  &52&4.58  &1   &3 &2.86&11 &10,15,5             &1,8,21    \\
Par 21     &7.60 &75  &-    &-   &-   &- &7.60  &75  &1 &1.50&6  &7                   &23        \\
V1685 Cyg  &1.18 &14  &0.16 &0.46&12  &52&1.02  &15  &6 &0.08&4  &15                  &8,3,4     \\
\vspace{0.2cm}
MWC 349    &7.70 &167 &3.6  &0.50&100 &20&10.4  &174 &11&1.55&4  &-                   &25,26        \\
LkH$\alpha$ 234
           &0.53 &107 &0.87 &0.65&43  &21&1.16  &149 &11&0.65&21 &10,15,13            &1,9,8,4   \\
PV Cep     &14.2 &77  &0.99 &0.50&141 &15&14.8  &76  &1 &0.41&1  &                    &19        \\
HD 200775  &0.94 &93  &0.69 &0.52&141 &21&1.23  &76  &4 &0.50&21 &10,15               &1,7,8     \\
AS 477     &1.10 &35  &0.49 &0.58&35  &34&0.84  &44  &5 &0.41&42 &10,15,13,-          &1,9,3,7,20\\
\vspace{0.2cm}
LkH$\alpha$ 233
           &11.1 &156 &0.60 &0.78&85  &37&11.6  &157 &2 &0.40&2  &10,15               &1,8       \\
HD 216629  &4.72 &105 &1.29 &0.67&80  &15&4.03  &112 &1 &0.20&1  &                    &27\\
MWC 1080   &2.00 &75  &2.18 &1.00&72  &13&0.42  &150 &10&1.00&30 &10,15,13,8.3        &1,9,8,28  \\
\hline
\end{tabular}
\label{tab3}
\vspace{0.2cm}

\underline{\textit{Notes:}}\\

\textit{GGD 18: Combined K-band degree of polarization and position angle are given here.}\\
\textit{MWC 349: The interstellar degree of polarization and position angle values are
taken from Yudin (1996) and are used to remove the interstellar contribution from observed polarization
values of this star.}\\

\underline{References}\\

1) Hillenbrand et~al. 1992; 2) Asselin, Manard and Bastien 1991; 3) Jain et~al. 1995; 
4) Oudmaijer et~al. 2001; 5) Whittet et~al. 1992; 6) Hutchinson et~al. 1994;
7) Breger 1974; 8) Vrba et~al. 1979; 9) Garrison and Anderson 1978; 10) Bastien 1982;
11) Bastien 1982; 12) Scarrott 1989; 13) Hall 1958; 14) Serkowski et~al. 1975; 15) This paper
16) Yudin and Evans 1998; 17) Barbier and Swings 1992; 18) Gnedin et~al. 1992;
19) Menard and Bastien 1992; 20) Vrba 1975; 21) Ward-Thompson et~al. 1985; 22) Marrco and
Forte 1978 23) Draper et~al. 1985; 24) Sato et~al. 1985; 25) Yudin 1994; 
26) Meyer et~al. 2001; 27) Heiles 2000; 28) Manset and Bastien 2001; 29) Bessell and Eggen 1972.\\
\end{table*}

\section {Data on polarization, outflows and binary companions of Herbig
Ae/Be stars}
\label{dat}
In recent years several studies on polarization, outflows and
binarity of Herbig Ae/Be stars have become available in the literature.
In Table~\ref{tab2} we list position angle of the binary companion
and outflow taken from the literature. The columns give, respectively, (1) object 
identification, 
(2-3) distance and spectral type taken from the literature, (4) secondary 
component position angle with
respect to primary measured from celestial north, (5) angular separation ($arcsec$) of the 
secondary from the primary star, (6) derived projected linear separation (AU) 
of the secondary component from 
the primary star, (7) the mode of detection, (8) outflow position angle, (9) references. 
In Table~\ref{tab3} we list polarization data on Herbig Ae/Be stars. The column 1 gives 
object identification, columns 2 and 3 give observed degree of polarization and position
angle. For most of the stars measurements at multiple epochs are available. Average values
are given here. 
Columns 4 and 6 give estimated contribution of interstellar polarization and its position
angle. Contributions to the observed polarization in young stellar objects are primarily
due to the scattering of stellar light by dust distributed in the cicumstellar environment
and that due to the interstellar medium. For most of the Herbig Ae/Be stars being considered here,
the observed polarization is dominated by the circumstellar component. The interstellar contribution
becomes relatively more important for stars which have smaller circumstellar polarization
either due to smaller quantities of scattering dust or due to unfavourable orientation, 
especially for more distant objects. 
We estimate the interstellar contribution to the polarization of the Herbig Ae/Be
stars by considering the observed polarization of normal stars, at different distances
in the direction of the object, from the catalogue Stellar polarization catalogs agglomeration 
by Heiles (2000). 
Around each Herbig Ae/Be star, a search is made for normal stars in circles of increasing angular radii.
Minimum of 10 stars are used to estimate the interstellar contribution to each target star.
Circles of radius 1$^{\circ}$, 2$^{\circ}$, 3$^{\circ}$, 5$^{\circ}$ and
7$^{\circ}$ are used to choose the stars. If less than 10 stars are found in 7$^{\circ}$,
those stars are used. Stoke's parameters U (=P$sin2\theta$) and Q (=P$cos2\theta$) are evaluated
from the degree of polarization (P) and position angle ($\theta$) for each star. U and Q parameters
thus evaluated are plotted against the distance to the respective stars. Stoke's parameters
U$_{ism}$ and Q$_{ism}$ representing interstellar polarization at the target star's distance 
are estimated by making a 
least-square fit. The interstellar polarization value  P$_{ism}$ and position angle 
$\theta_{ism}$ are calculated as\\

P$_{ism}$ =$ \sqrt{(U_{ism})^{2}+(Q_{ism})^{2}}$ \\ 

$\theta_{ism} =(1/2)  tan^{-1}(U_{ism}/Q_{ism})$     \\  

P$_{ism}$ and $\theta_{ism}$, and probable errors $\epsilon_{p_{ism}}$ and $\epsilon_{\theta_{ism}}$
in their estimation,
are given in columns 4, 6 and 5, 7 of Table~\ref{tab3} respectively.
Stoke's parameters U$_{\star}$ and  Q$_{\star}$ representing the observed polarization for the target  
stars are evaluated from the observed degree of
polarization and position angle measured at each epoch. The Stoke's
parameter U$_{i}$ and  Q$_{i}$ representing the intrinsic (circumstellar polarization of 
the target star) polarization are estimated as\\

U$_{i} = U_{\star}-U_{ism}$\\

Q$_{i} = Q_{\star}-Q_{ism}$\\

The intrinsic polarization P$_{i}$ and position angle $\theta_{i}$ are then evaluated as\\

P$_{i}$ =$ \sqrt{(U_{i})^{2}+(Q_{i})^{2}}$\\

$\theta_{i} =(1/2)  tan^{-1}(U_{i}/Q_{i})$ \\
      
In the case of V892 Tau,
HD 150193 and Par 21 not many stars are available to determine the interstellar polarization 
and position angle. However, we note that they have relatively high observed polarization 
and are
at relatively smaller distances. With relatively small intrinsic interstellar contribution, the
observed polarization for the stars is considered to represent their 
intrinsic polarization in this study.
Columns 8 and 9 give average values of intrinsic degree of polarization and position 
angle for Herbig Ae/Be stars corrected for the estimated interstellar contribution.
Column 10 gives number of multiple epoch observations used for each star.
Columns 11 and 12 give the probable errors in the polarization and 
position angle values inclusive of the dispersions in the individual
measurements at different epochs. The median value of 
dispersion in position angle is $\approx 8 ^{\circ}$. 
Only for some stars namely, T Ori, HD 144432, AS 477 and MWC 1080, the 
dispersion in position angles is  
found to be more than or equal to 30$^{\circ}$. Column 13 gives various apertures used in the polarimetric
measurements and  column 14 gives the references.\\

 \begin{table}
\caption[]{Relative orientations of binary, polarization and
outflow position angles for Herbig Ae/Be stars.}
\begin{tabular}{l c c c}
\hline\hline
Object&$\vert{\Delta_{po}-90}\vert$&$\vert{\Delta_{bp}-90}\vert$&$\vert{\Delta_{bo}-90}\vert$\\
      &($^{\circ}$)                &($^{\circ}$)                &($^{\circ}$)    \\
\hline 
V633 Cas       &61      &84  &67  \\
V376 Cas       &6       &-   &-   \\
XY Per         &-       &41  &-   \\
V892 Tau       &-       &70  &-   \\
\vspace{0.2cm}
UX Ori         &-       &73  &-    \\
HD 35187       &-       &50  &-     \\
CO Ori         &-       &25  &-   \\
HK Ori         &89      &29  &28  \\
T Ori          &-       &63  &-   \\
\vspace{0.2cm}
V380 Ori       &55,32   &23  &58,35\\
LkH$\alpha$ 208 &65     &1   &24   \\
GGD 18         &21      &83  &14   \\
R Mon          &1       &30  &61   \\
GU CMa         &-       &70  &-    \\
\vspace{0.2cm}
Z CMa          &23      &40  &27   \\
HD 53367       &-       &13  &-    \\
NX Pup         &-       &78  &-   \\
HD 76534       &-       &81  &-   \\
Her 4636       &76      &38  &24    \\
\vspace{0.2cm}
HD 141569      &-       &47  &-   \\
HD 144432      &-       &88  &-   \\
HR 5999        &-       &84  &-   \\
HD 150193      &-       &87  &-   \\
KK Oph         &-       &23  &-   \\
\vspace{0.2cm}
HD 163296      &28      &-   &-   \\
AS 310         &-       &-   &8   \\
R CrA          &31      &-   &-     \\
T CrA          &42      &4   &52    \\
Par 21         &10      &-   &-     \\
\vspace{0.2cm}
V1685 Cyg      &-       &82  &-   \\
MWC 349         &74      &16  &0     \\
LkH$\alpha$ 234  &13,13  &76  &27,1   \\
PV Cep         &2       &-   &-   \\
HD 200775      &84      &2   &4   \\
\vspace{0.2cm}
AS 477         &22      &89  &21   \\
LkH$\alpha$ 233 &17,23   &-   &-    \\
HD 216629      &-       &55  &-     \\
MWC 1080       &0       &26  &64    \\
\hline

\end{tabular}
\label{tab4}
\end{table}

\section {Relationship between polarization angles, outflow directions and
binary orbital planes}
\label{rel}

Various alignments between the linear polarization vector and
other phenomena have earlier been looked for by several authors.
The polarization vector is found to be perpendicular to the
optical jets or CO molecular outflows for most sources
(Mundt and Fried, 1983; Nagata, Sato and
Kobayashi, 1983; Hodapp, 1984; Bastien, 1987). Bastien (1987)
compiled a list of 23 young outflow sources for which the central
source  has been found and measured its linear polarization.
The list included 8 TTS, 8 HAeBe, 1FUOR, 4 YSO, and 2 objects
with unknown status. The distribution of the difference of the
outflow position angle, $\theta_{jet}$, and the linear
polarization position angle, $\theta_{*}$, was studied and it was
found that in $61\%$ of the sources these two directions are
perpendicular to each other to within $30^{\circ}$. This was done
to distinguish between two models of the origin of polarization namely
oblate configuration as in a circumstellar disk around the star 
and prolate configuration as in two oppositely directed jets. 
Both the models have the same $sin^{2}i$ dependence, and hence it is difficult to
distingush between elongated and flat models. However, the polarization is usually 
perpendicular to
the scattering plane. One expects the polarization vector to be along the axis for
flat models and perpendicular to it for the elongated distribution. Thus for an outflow source,
the polarization vector and the outflow position angle would be roughly parallel to each other in a flat
model and perpendicular to each other for an elongated model. Also, if the source is 
in a binary system and if the
disk as well as the orbital plane are coplanar, the polarization position angle would be 
roughly perpendicular to the binary position angle in a flat model and parallel to the binary 
position angle in an 
elongated model (for favoured orientations). In the following we present a study of 
correlation between the binary, polarization and outflow position angles in Herbig Ae/Be stars.\\

\begin{figure*}[!]
\includegraphics*{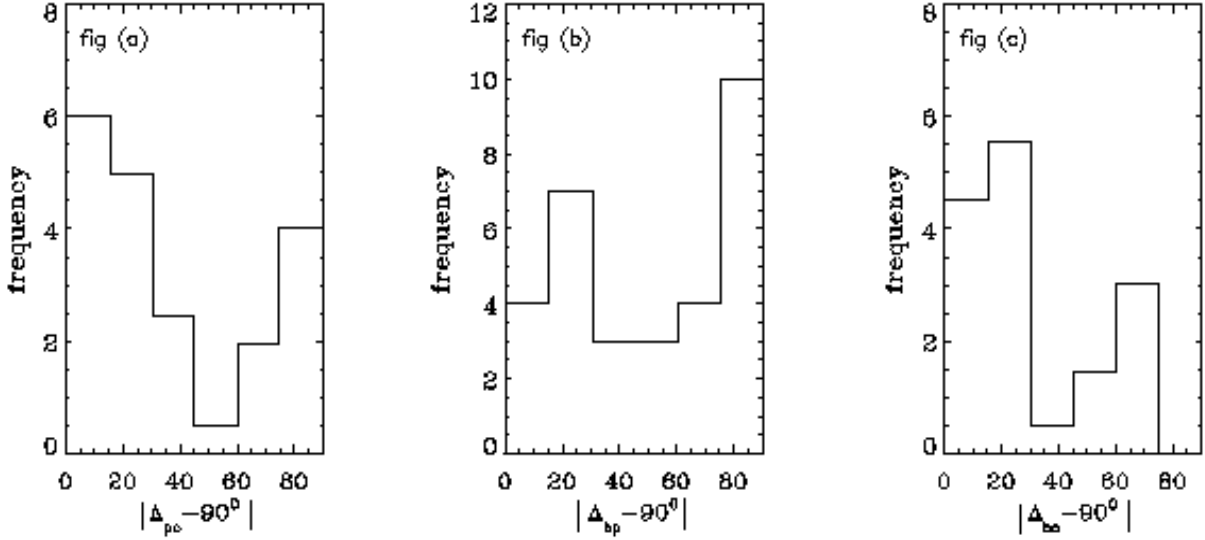}
\caption{Figure (a), the frequency distribution of the difference in polarization
and outflow position angles. Figure (b), the frequency distribution of the difference 
in polarization and binary position angles. Figure (c), the frequency distribution of 
the difference in binary and outflow position angles. }
\end{figure*}

In Table~\ref{tab4}, the relative orientations of binary, polarization and
outflow position angles are presented for individual stars.
Columns in Table~\ref{tab4} give respectively, (1) object identification, (2) $\vert{\Delta_{po} - 90}\vert$,
where $\Delta_{po} = {\theta_{p}-\theta_{o} \pm n\pi}$, (3) $\vert{\Delta_{bp} - 
90}\vert$, where
$\Delta_{bp} = {\theta_{b} - \theta_{p} \pm n\pi}$, (4) $\vert{\Delta_{bo} - 90}\vert$, 
where $\Delta_{bo} = {\theta_{b}-\theta_{o}\pm n\pi}$. $\theta_{b}, \theta_{p}, 
\theta_{o}$ 
are the observed binary, polarization and outflow position angles respectively and $\pi$ stands for
$180^{\circ}$. In Table~\ref{tab4},
17 ($\approx 85\%$) of the 20 outflow sources have outflow  position angle within $30^{\circ}$ 
of being either perpendicular or parallel to the
polarization position angle. Histogram (a) shown in Figure 1 gives the 
frequency distribution of the difference in polarization and outflow position angles.
Sources with outflow position angle within $30^{\circ}$ of being perpendicular
to polarization position angle ($\vert{\Delta_{po}-90}\vert$ $\leq$ $30^{\circ}$) are
found to be 55\%. Polarization position angle can be perpendicular to outflow
if the circumstellar disk is optically thick and the outflow is perpendicular to it. As can be 
seen from Table 3, six sources,
namely, V376 Cas, GGD 18, R Mon, Par 21, PV Cep and LkH$\alpha$ 233
with polarization position angle
perpendicular to outflow direction (within $30^{\circ}$) have polarization 
value greater than 3\%. This can happen when highly polarized scattered light from
the optically thin polar regions superimposed on strongly attenuated unpolarized
direct light from the central star reaches the observer whose line of sight is close to being
edge-on. Five sources given in Table 3, namely, Z CMa, HD 163296, LkH$\alpha$ 234, AS 477
and MWC 1080 with polarization position angle perpendicular to the outflow ( within $30^{\circ}$)
have polarization value less than 3\%. This situation can arise when polarized scattered
light from the optically thin polar regions superimposed on weakly attenuated direct 
light from the central star reaches the observer whose line of sight is relatively away from 
the equatorial region. From Table~\ref{tab4}, the sources with outflow position angle 
within $30^{\circ}$ of being
parallel to polarization position angle ($\vert{\Delta_{po}-90}\vert$ $\geq$ $60^{\circ}$)
are found to be 30\%. Such a situation can arise
when star has an oblate envelope or an optically thin disk. Sources satisfying this
condition namely, V633 Cas, HK Ori, LkH$\alpha$ 208, Her 4636 and HD 200775 have 
polarization values less than 3\% except for MWC 349.
Polarization values larger
than $\approx 2\%$ are hard to obtain owing to the presence of direct unpolarized light
from the star and probably also because extremely oblate envelope do not occur 
(McLean \& Brown, 1978 ).
However, we note here that the interpretation in terms of optically thin dust distribution
may not be correct for some of these stars. Three of these stars (HK Ori, LkH$\alpha$ 208 and
HD 200775) were also studied by Hillenbrand et al. (1992) and were found to have Group I SEDs
indicative of optically thick disks. Scattering off the surface of the optically thick disk
(Whitney \& Hartmann, 1992) may be responsible for the observed polarization for these objects.
Among the 11 sources with polarization vectors perpendicular to the
outflow, 7 are in common with Hillenbrand et al. (1992). Four of these (V376 Cas, R Mon, Par 21 and
LkH$\alpha$ 233) show Group II SEDs indicative of optically thick disks surrounded by 
extended envelopes, and 3 (HD 163296, LkH$\alpha$ 234 and MWC 1080) show Group I SEDs.
 Three sources namely, V380 Ori, R CrA and T CrA have
outflow position angle neither parallel nor perpendicular to the polarization
position angle. This can happen when the outflow direction is not perpendicular to
the circumstellar/circumbinary optically thin/thick disk. Fendt \& Zinnecker (2000)
showed that in some cases, protostellar jets and counter jets are misaligned and
the reason they suggest for the misalignment is the bending of jets due to the motion
of jet source in a binary system. Similar effects could be responsible for the outflow being
neither perpendicular nor parallel to the polarization position angle in the binary
systems V380 Ori and T CrA. R CrA is suggested to have an interstellar disk (Ward-Thompson et al. 1985)
around it which 
bends the outflow originating from it. V380 Ori, LkH$\alpha$ 234 and LkH$\alpha$ 233 have
outflows associated with them in two directions. Each of them is counted 1/2 in the histogram.
A two sided Kolmogorov-Smirnov test shows that the frequency distribution shown in Figure (a) is different 
from a random distribution to 97\%.
In Figure 2 the difference in polarization and outflow position angles
($\vert{\Delta_{po}-90}\vert$) is plotted against intrinsic polarization P$_{i}$.
Error bars shown in Figure 2 result from probable errors in polarization measurements,
uncertainties in the estimation for interstellar polarization, dispersions in multi-epoch
values and the uncertainties in the outflow position angles. The probable errors in the outflow 
position angles are typically 5-20$^{\circ}$ (Bastien 1987) depending on the method of outflow detection
(optical/NIR, radio) but are not given explicitly by the authors for most of the sources.
Where not available, we have taken a value 5$^{\circ}$ for optical/NIR and 20$^{\circ}$ for radio observation. 
From Figure 2 it can be seen that objects
with $\vert{\Delta_{po}-90}\vert$ $\leq$ $30^{\circ}$ tend to exhibit large values of
polarization indicative of optically thick disks with extended envelopes. These results strengthen the
correlation found by Bastien (1987).\\

\begin{figure*}[!t]
\includegraphics*{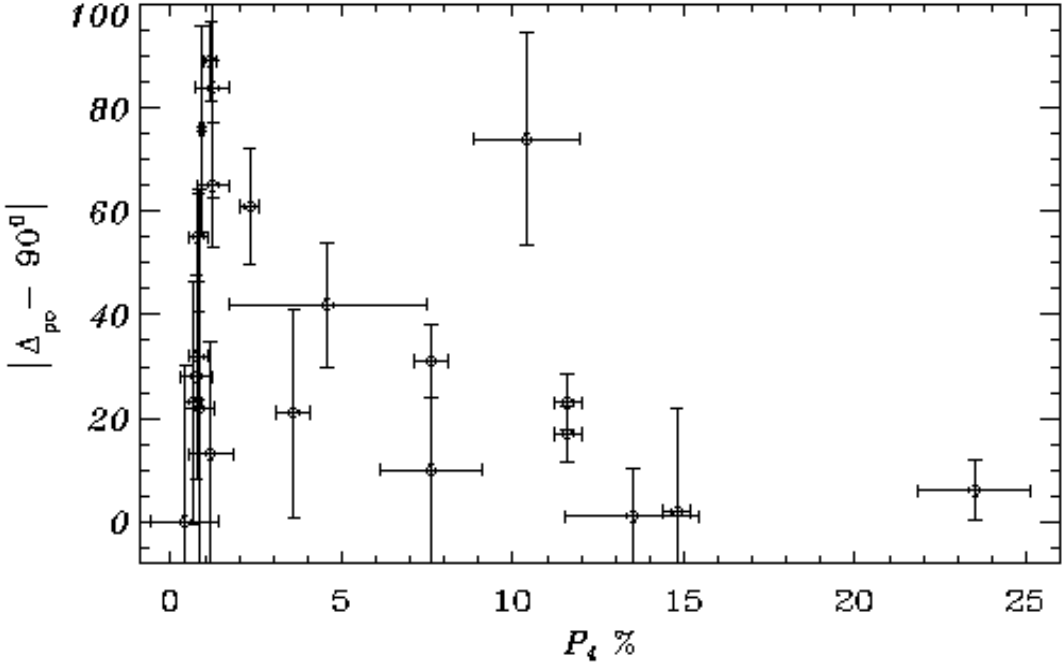}
\caption{Difference in polarization and outflow position angles plotted against
intrinsic polarization P$_{i}$.}
\label{figure 2}
\end{figure*}

Histogram (b) in Figure 1 shows the frequency 
distribution of difference in polarization position angle and binary position angle.
In Table~\ref{tab4}, 
25 ($\approx 81\%$) out of 31 binary systems have polarization position angle within
$30^{\circ}$ of being either perpendicular or parallel to the binary position angle.
Among 31 sources, 10 ($\approx 32\%$) have polarization position angle within 15$^{\circ}$
of being parallel to the binary position angle while 45\% of the sources show the 
two position angles parallel to within $30^{\circ}$. Parallelity of the two position angles 
can result if the binary component is coplanar with an optically
thick disk since polarization arises due to the scattering of stellar light by the
dust distributed in the polar regions.
In 35\% of the sources given in Table~\ref{tab4} polarization position angle is
perpendicular to the binary position angle within $30^{\circ}$. This situation can arise when the
binary component is coplanar with an optically thin disk where polarization arises due to the
scattering of stellar light by the dust distributed in equatorial region.
Alternatively, polarization could be caused by scattering off the surface of an optically
thick disk (Whitney \& Hartmann, 1992).
A two sided Kolmogorov-Smirnov test shows that the frequency distribution 
shown in Figure (b) is different from a random distribution to 84\%.
However, the binary component position angle does not in a strict sense
represent the orbital plane (except in edge-on systems) owing to
the projection effects and this will weaken the correlation.
We find here that there exists a correlation 
between binary  position angle and polarization position angle
inspite of the projection effect. This indicates that in actual case there exist a
correlation that is even stronger than observed.\\

Histogram (c) in Figure 1 shows the frequency 
distribution of difference in binary and outflow position angles. Among binary systems, 
there are 15
sources which are associated with outflows also. It can be seen that 10 ($\approx 67\%$) of the 15 
binary sources have the binary position angle within $30^{\circ}$ of being   
perpendicular to outflow position angle.
A two sided Kolmogorov-Smirnov test shows that the frequency distribution shown 
in Figure (c) is different from a random distribution to 96\%.
Here again it should be noted that the projection effects 
on the binary orbital planes would weaken even a perfect correlation (perpendicularity) 
between the outflow and binary position angles. So the observed correlation is 
relatively more significant.\\

The results presented in Figure 1, the various correlations and their interpretation 
discussed above must be viewed with caution as they are based on relatively small number statistics
(17/20, 25/31, 10/15). In particular, for the restricted sample of binary sources with outflows, only 
7($\approx$ 50\%) are compatible with the interpretation of their polarization position angle
relative to the outflow and binary position angles. Also, projection effects on binary orbital
planes, as noted earlier, tend to reduce the observed correlations.
Correlations between the different position angle differences and binary component 
separation were also investigated. For $\vert\Delta_{po} - 90\vert$,  $\vert\Delta_{bp} - 90\vert$
and $\vert\Delta_{bo} - 90\vert$ as a function of log (projected linear separation in AU) we obtain
very low linear correlation coefficients (0.26, 0.17, 0.20 respectively) indicating poor correlation.

\section{Discussion}
\label{dis}

The geometrical relationships between binary position angle, polarization
position angle and outflow position angle studied here can be compared with
those expected from different mechanisms of formation of binary system,
circumstellar disks and young stellar objects. The favoured binary
formation mechanisms are: (1) capture process - in which two independently
formed stars can be captured into orbits under certain conditions (Hills 1976;
Boss 1988; Hills \& Day 1976; Mansbach 1970); (2) fission process -  
in which as a star contracts
towards mainsequence it spins up and the ratio of rotational to
gravitational energy ($\beta$) increases.   When ($\beta$) increases beyond a
certain critical value, the star becomes unstable to non-axisymmetric
perturbations. It has been hypothesized that breakup into orbiting
subcondensations then occurs 
(Ruzmaikina 1981a,b; Williams \& Tohline 1988); (3) fragmentation - 
in which a cloud that is initially differentially rotating (Myhill \& Kaula 1991),
or has a milder exponential-type density profile (Boss 1991), 
fragments. Another mode of fragmentation was suggested by Zinnecker (1990)
where he proposed that since most of the interstellar clouds often show an
elongated, filamentary structure, it rotates about an axis perpendicular to the
cylindrical axis. After fragmentation of the cylinder (Bastein 1983) the 
fragments move towards each other along the cylindrical axis until they reach
a Keplerian orbit. This mechanism results in the formation of binary systems with
wide separation; (4) disk fragmentation - in which a relatively slowly rotating 
protostar collapses through the adiabatic phase without fragmentation and will 
form a disk-like structure. Equilibrium Keplerian disks around central stars have
the possibility of fragmenting due to gravitational instabilities (Adams,
Ruden \& Shu 1989; Shu et al. 1990). In all the above mentioned binary
formation mechanisms except the capture, one would expect the disk around
each component or the circumbinary disk and the binary orbit
to be coplanar. Results presented
in this paper support those binary formation mechanisms in which disk around
each component or circumbinary disk and the binary orbital plane are 
coplanar. 
While polarization position angle could be either parallel or perpendicular
to the circumstellar disk depending on the optical depth
and the presence of extended envelopes, the outflow direction is 
expected to be always perpendicular to the disk. If the disks are coplanar with
binary orbital plane, then the outflow position angle should be perpendicular to the 
binary position angle except for any projection effect. From Figure 3 it is seen 
that out of 15 outflows listed here 10 of them have difference in
binary component and outflow to be within $30^{\circ}$ of being perpendicular to each other.
The polarimetric observations used in this work have genarally been made 
with large apertures ($\approx 10-15  arcsec$) encompassing both the components of 
the Herbig Ae/Be binaries (with typical angular separation of
$\leq 2 arcsec$). Since the optical light is dominated by the primary component the 
observed polarization is determined by the distribution of scattering matter around it. The outflow
in a binary Herbig Ae/Be star could be driven by either the primary or even the 
secondary which is generally a low mass infrared object, having its own circumstellar disk. The 
results on the correlations between the various position angles discussed above therefore indicate
coplanarity of the disks around the individual components. For low mass T Tauri binaries, polarimetry
on individual component stars (Jensen et~al. 2000; Wolf et~al. 2001 and Monin et~al. 2000) have
shown that in a majority of the binary systems the linear polarization vectors from individual
stars are within $30^{\circ}$ of being parallel. Resolved polarimetric measurements of the Herbig Ae/Be binary
components will be needed to confirm the results presented here.\\

\section {Conclusions}
\label{con}

Circumstellar matter around Herbig Ae/Be stars causes polarization in the 
light from these objects due to scattering from dust. The position angle 
of the observed polarization depends on whether the scattering is dominated by 
optically thin disk or by optically thin/thick scattering dust distributed perpendicularly
to an optically thick disk. Bipolar outflows are generally constrained to be
perpendicular to the disks. In binary Herbig Ae/Be stars, the binary orbital 
plane may also be correlated with the polarization and outflow directions if 
the binary formation mechanism derives the binary orbital angular momentum
and the individual star's circumstellar disk from the angular momentum of 
the same rotating clouds. In this paper we have studied the observed correlation
between the different position angles. Our results can be summarized as follows:\\

1) Out of 20 outflow sources, 17 sources ($\approx 85\%$) have the outflow  
position angle within $30^{\circ}$ of being either parallel or
perpendicular to the polarization position angle.\\

2) In 25 ($\approx 81\%$) out of 31 sources, the direction of binary 
position angle is within $30^{\circ}$ of being either parallel or perpendicular
to the polarization position angle.\\

3) In 10 ($\approx 67\%$) out of 15 outflow sources, the binary position angle is within 
$30^{\circ}$ of being perpendicular to the outflow position angle.\\

These results are consistent with binary star formation senarios in which the 
circumstellar disk planes are parallel to the binary orbital plane.
However, it must be noted that these results are subject to small number statistics
as they are based on samples that are small and need to be enlarged.\\

\begin{acknowledgement}
We thank the referee for valuable comments that have led to substantial improvements in 
the analysis and presentation of the paper.
\end{acknowledgement}

\end{document}